\documentclass[prl,showpacs,superscriptaddress, twocolumn]{revtex4}
\usepackage{dcolumn}
\usepackage{graphicx}

\begin{document}

\raggedbottom

\title{Deceleration and electrostatic trapping of OH radicals}

\author{Sebastiaan Y.T. van de Meerakker}
\affiliation{Fritz-Haber-Institut der Max-Planck-Gesellschaft,
Faradayweg 4-6, 14195 Berlin, Germany} \affiliation{FOM-Institute
for Plasmaphysics Rijnhuizen, Edisonbaan 14, 3439 MN Nieuwegein,
The Netherlands}

\author{Paul H.M. Smeets}
\affiliation{FOM-Institute for Plasmaphysics Rijnhuizen, Edisonbaan 14,
3439 MN Nieuwegein, The Netherlands}

\author{Nicolas Vanhaecke}
\affiliation{Fritz-Haber-Institut der Max-Planck-Gesellschaft,
Faradayweg 4-6, 14195 Berlin, Germany}

\author{Rienk T. Jongma}
\affiliation{Space Research Organization Netherlands, Sorbonnelaan 2,
3584 CA Utrecht, The Netherlands}

\author{Gerard Meijer}
\affiliation{Fritz-Haber-Institut der Max-Planck-Gesellschaft,
Faradayweg 4-6, 14195 Berlin, Germany}

\date{\today}

\begin{abstract}
A pulsed beam of ground state OH radicals is slowed down using a Stark
decelerator and is subsequently loaded into an electrostatic trap.
Characterization of the molecular beam production, deceleration and trap
loading process is performed via laser induced fluorescence detection
inside the quadrupole trap. Depending on details of the trap loading
sequence, typically $10^5$ OH ($X^2\Pi_{3/2}, J=3/2$) radicals are trapped
at a density of around $10^7$ cm$^{-3}$ and at temperatures in the
50-500~mK range. The 1/e trap lifetime is around 1.0~second.
\end{abstract}


\maketitle

Getting ever better control over both the internal and external degrees of
freedom of gas-phase molecules has been an important theme in molecular
physics during the last decades. This control is important, for instance,
in spectroscopic investigations, in collision and reactive scattering
experiments and in photo-dissociation, or half-collision, studies. Seeded
molecular beams, both continuous and pulsed, have been extensively used
to produce samples of molecules in only the lowest internal energy states,
with typical rotational temperatures of a few K \cite{Scoles1988+1992}.
In these molecular beams the longitudinal velocity distribution is also
well defined, and translational temperatures of around 1~K are
obtained in the moving frame of the molecular beam. Further control over
the molecules is achieved when their orientation in space is actively
manipulated, using either static electro-magnetic fields and/or radiation
fields \cite{Stapelfeldt2003}. Sophisticated and powerful detection schemes
have been developed to experimentally study (half-) collisions
\cite{Sato2001,Rakitzis2004} and reactions \cite{Liu2001+2003} of the
thus prepared molecules in the required detail.

Over the last years our group has been developing methods to get improved
control over the absolute velocity and over the velocity spread of
molecules in a molecular beam \cite{Bethlem2003}. These methods rely on the,
quantum-state specific, force that polar molecules experience in electric
fields. This force is rather weak, typically some eight to ten orders of
magnitude weaker than the force that the corresponding molecular ion
experiences in the same electric field. This force nevertheless suffices
to achieve complete control over the motion of polar molecules, using
techniques akin to those used for the control of charged particles.
We have explicitly demonstrated this by the construction of
two types of linear accelerators \cite{Bethlem1999+2002b}, a
buncher \cite{Crompvoets2002}, a trap \cite{Bethlem2000} and a storage
ring \cite{Crompvoets2001} for neutral polar molecules. Using these tools,
state-selected molecular beams with a computer-controlled absolute
velocity and with record-low (longitudinal) temperatures have been produced.
This holds great promise for the use of decelerated molecular beams
in metrology, i.e., in experiments aimed at testing fundamental
symmetries \cite{Tarbutt2004}.

To be able to exploit the possibilities that these new tools
offer for collision and reactive scattering experiments, the
fraction of the pulsed molecular beam that is decelerated and/or
trapped needs to come closer to unity, i.e., the 6-dimensional
phase space acceptance of the various elements needs to be
increased to better match to the typical emittance of a molecular
beam. In addition, deceleration and trapping needs to be
performed on those molecules that are of most relevance in
collision and reactive scattering experiments; thus far,
electrostatic trapping after Stark deceleration has only been
demonstrated for ND$_3$ molecules \cite{Bethlem2000,Bethlem2002a}.

We here report the deceleration and electrostatic trapping of ground state
OH ($X^2\Pi_{3/2}, J=3/2$) radicals. The experiments are performed in a
new generation molecular beam deceleration machine, designed such that a
large fraction of the molecular beam pulse can be slowed down and trapped.
The role of the omnipresent OH radical as intermediate in many chemical
reactions has made this a benchmark molecule
in collision and reactive scattering studies. Hexapole state-selection and
focusing, e.g., transverse phase space manipulation, of a beam of OH is well
documented and often used in these studies \cite{Beek2001}. The possibility
to manipulate the longitudinal phase-space distribution of OH radicals inside
a decelerator has recently been demonstrated as well \cite{Bochinsky2003}.

\begin{figure}
    \centering
    \resizebox{\linewidth}{!}
    {\includegraphics[-97,117][712,672]{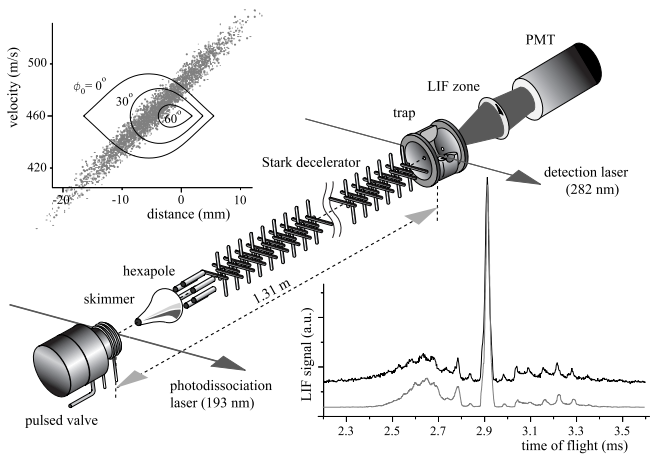}}
    \caption{Scheme of the experimental setup. The longitudinal phase space
    acceptance of the decelerator for OH ($J=3/2, |M_J\Omega|=9/4$)
    is given for three different phase angles $\phi_0$, together with the longitudinal
    phase space distribution of the OH beam at the entrance of the decelerator
    (shaded area). The observed LIF signal as a function of time after OH production
    with the decelerator operating at $\phi_0 = 0^{\circ}$ (upper curve) is shown
    together with a simulation (lower curve).}
    \label{fig:Setup}
\end{figure}

The experimental setup is schematically shown in
Fig.~\ref{fig:Setup}. A pulsed beam of OH with a mean velocity of
460~m/s and with a velocity spread of 15~\% (full width half
maximum, FWHM) is produced via ArF-laser dissociation of HNO$_3$
seeded in Kr near the orifice of a pulsed valve. Production thus
takes place at a well-defined time and at a well-defined
position, which is ideal for coupling the molecular beam into the
decelerator. More than 95~\% of the OH radicals in the beam
reside in the lowest rotational and vibrational level of the
$X^2\Pi_{3/2}$ spin-orbit manifold of the electronic ground
state. This population is distributed over the two
$\Lambda$-doublet components of the $J=3/2$ level, which has a
zero-field splitting of 0.055~cm$^{-1}$. Only OH radicals in
low-field seeking quantum states, i.e., radicals in the upper
$\Lambda$-doublet component which splits into $|M_J\Omega|=3/4$
and $|M_J\Omega|=9/4$ components in an electric field, are of
relevance in this experiment. The force that the OH radicals
experience in electric fields is, in the linear regime,
proportional to $|M_J\Omega|$. Only molecules in (both $|M_F|$
hyperfine sublevels of) the $J=3/2, |M_J\Omega|=9/4$ component
are therefore decelerated and trapped in the present experiments,
although molecules in both $|M_J\Omega|$ components contribute to
the laser induced fluorescence (LIF) signal of the
non-decelerated beam that is passing through the trap.

After passage through a 2~mm diameter skimmer, the molecular beam enters the
deceleration chamber and is focussed into the Stark decelerator by a short
hexapole. The hexapole matches the transverse phase space distribution of the
OH ($J=3/2, |M_J\Omega|=9/4$) radicals at the entrance of the decelerator to
the transverse acceptance of the decelerator. A detailed description of the
operation principle of a Stark decelerator and an electrostatic quadrupole trap,
as well as of (the importance of) phase-space matching between the various
elements in the molecular decelerator beam-line is given elsewhere \cite{Bethlem2002a}.
In the current design the dimensions of the decelerator are scaled up
by a factor of two relative to earlier designs. With the electric field stages in
the decelerator now at a distance of 11~mm, and with a $4 \times 4$ mm$^2$ transverse
acceptance area between the two 6~mm diameter parallel electrodes that make up one
electric field stage, the total spatial acceptance of the decelerator has been
increased by a factor eight. This upscaling has been performed while maintaining
the same electric field strength inside the decelerator; in the present experiments
a voltage difference of 40~kV is switched on and off in an electric field stage.
The area within the outer contour in the upper left corner in
Fig.~\ref{fig:Setup} is the thus obtained longitudinal acceptance of the decelerator
for phase-stable transport of OH ($J=3/2, |M_J\Omega|=9/4$) without deceleration.
With a distance from the source to the decelerator of only about 10~cm, this longitudinal
acceptance largely encompasses the longitudinal emittance of the pulsed OH beam,
indicated by the shaded area. When operating the decelerator at $\phi_0=0^{\circ}$
for a synchronous molecule with a velocity of 450~m/s, a time of flight (TOF) profile
of the OH ($J=3/2$) radicals exiting the decelerator as shown in the lower right
corner of Fig.~\ref{fig:Setup} is observed. The main portion of the OH beam is
transported through the machine as a compact package (less than 25~$\mu s$
FWHM in the TOF profile), independent of the length of the decelerator. The TOF profile
resulting from a three-dimensional trajectory calculation, shown underneath the
measurement, quantitatively reproduces the observation. From the calculations, the
contributions of the individual $|M_J\Omega|$ components to the TOF profile can be
identified. In addition, these calculations enable a detailed understanding of the
untrapped dynamics in the decelerator, which manifests itself by the features in the
wings of the observed TOF profile.

For deceleration of the molecular beam, operation at a phase angle
0$^{\circ} < \phi_0 < $90$^{\circ}$ is required. As the longitudinal acceptance of
the decelerator rapidly decreases with increasing phase angle it is often advantageous
to operate the decelerator at a relatively low phase angle, i.e., to extract a smaller
amount of energy from the molecular beam per deceleration stage, and to use more
deceleration stages instead. The decelerator used in the current experiment consists
out of 108~deceleration stages, and has a length of 1188~mm. The large number of
deceleration stages also enables the deceleration of molecular beams with a higher
initial velocity; in the present experiment this enables the use of Kr for seeding
instead of Xe, which reduces cluster formation in the expansion. In experiments
that we have performed to produce decelerated beams of metastable NH radicals,
it was unfavorable to use Xe as a carrier gas as it efficiently quenches the
electronically excited species. Only by performing 266~nm dissociation of HN$_3$
seeded in Kr a sufficiently intense beam of NH ($a^1\Delta, J=2$) radicals could be
produced, which we have subsequently decelerated from 550~m/s to 330~m/s (data not shown).

The electrostatic trap consists of a ring electrode, centered
21~mm downstream from the last electrodes of the decelerator,
with an inner radius $R$ of 10~mm and two hyperbolic end-caps
with a half-spacing of $R/\sqrt{2}$. This trap is scaled up by a
factor two compared to the trap that we have used thus far
\cite{Bethlem2002a}. The molecular beam enters the trap through a
4~mm diameter hole in the first end-cap. This hole is the only
vacuum connection between the deceleration chamber and the trap
chamber, and the pressure in the trap chamber can be kept at 5
$\times$ 10$^{-9}$ mbar under operating conditions. Optical
access to the trap is provided by two 6~mm diameter holes in the
ring electrode. The beam of a frequency doubled pulsed dye laser
is sent through these holes to perform LIF detection of the OH
radicals at the center of the trap. Excitation is performed on
the $Q_1(1)$ transition of the $A ^2\Sigma^+, v=1 \leftarrow
X^2\Pi_{3/2}, v=0$ band around 282~nm, selectively detecting OH
($J=3/2$) radicals in the upper $\Lambda$-doublet component.
Off-resonant fluorescence on the $A ^2\Sigma^+, v=1 \rightarrow
X^2\Pi, v=1$ band around 313~nm is imaged through a 6~mm diameter
opening in the second end-cap onto a photomultiplier tube (PMT).
Stray light from the laser is largely avoided by using light
baffles and is suppressed by optical filtering in front of the
PMT.

\begin{figure}
    \centering
    \resizebox{\linewidth}{!}
    {\includegraphics[73,75][536,519]{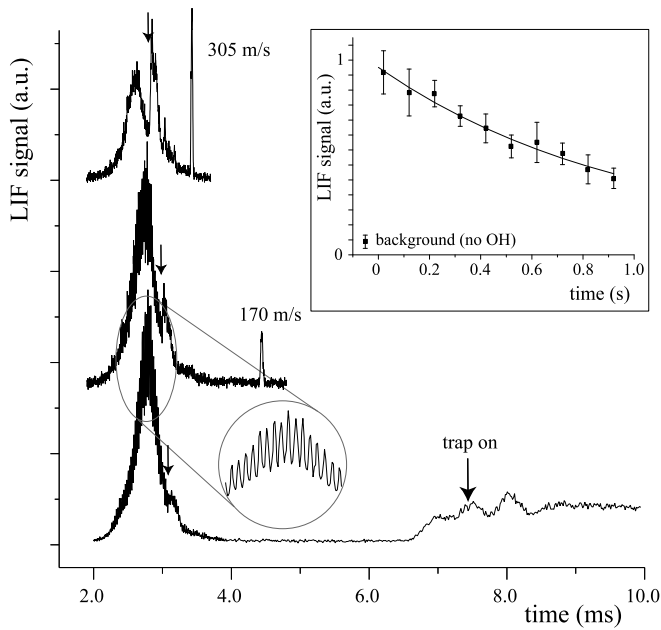}}
    \caption{LIF signal of OH ($J=3/2$) radicals at the center of the quadrupole
    trap as a function of time after OH production for two different deceleration
    sequences (upper two curves) and for a deceleration and trapping sequence (lower
    curve). In the inset, the LIF signal intensity of the electrostatically
    trapped OH radicals is shown on a longer time-scale.}
    \label{fig:TOF}
\end{figure}

In Fig.~\ref{fig:TOF} the intensity of the LIF signal of OH
($J=3/2$) radicals at the center of the trap is shown as a
function of time after firing the dissociation laser, using
different deceleration and trap loading sequences. In the
measurement shown in the upper curve, the decelerator is operated
at a phase angle of $\phi_0 = 50^{\circ}$, extracting about
0.9~cm$^{-1}$ of kinetic energy from the synchronous molecule in
every deceleration stage; OH radicals with an initial velocity of
around 465~m/s are decelerated to a final velocity of around
305~m/s. Due to a better spatial confinement, the peak density of
the decelerated portion of the molecular beam is higher than the
peak density of the non-decelerated beam that passes through the
trap. The hole in the TOF profile of the fast beam due to the
disappearance of OH radicals that are decelerated is indicated by
an arrow. In the measurement shown in the middle curve, OH
radicals with an initial velocity around 440~m/s are decelerated
to 170~m/s, operating the decelerator at a phase angle of $\phi_0
= 70^{\circ}$. In the lower curve, the observed TOF profile is
shown when the decelerator operates at $\phi_0=77^{\circ}$,
extracting about 1.2~cm$^{-1}$ of kinetic energy per deceleration
stage. OH radicals with an initial velocity of 428~m/s ($E_{kin}$
= 130~cm$^{-1}$) exit the decelerator with a velocity of around
21~m/s. The OH radicals in the non-decelerated part of the beam
also experience the switched electric fields inside the
decelerator. This leads to a highly structured phase-space
distribution and results in rich oscillatory structure on the TOF
profile of the fast beam, as shown enlarged in the figure.  The
slow beam of OH radicals is loaded into the electrostatic trap
with voltages of 7~kV, 15~kV and -15~kV on the first end-cap, the
ring electrode and the second end-cap, respectively. This creates
a potential hill in the trap that is higher than the remaining
kinetic energy of the molecules. The OH radicals therefore come
to a standstill near the center of the trap, around 7.4~ms after
their production. At that time, indicated by an arrow in
Fig.~\ref{fig:TOF}, the first end-cap is switched from 7~kV to
-15~kV to create a (nearly) symmetric 500~mK deep potential well.
After some initial oscillations, a steady LIF signal is observed
from the OH radicals in the trap. The LIF intensity of the
trapped OH radicals is about 20~\% of the peak LIF intensity of
the non-decelerated beam of OH that passes through the trap. The
observed LIF signal of the trapped OH radicals corresponds to a
total number of about $10^5$ OH ($J=3/2, |M_J\Omega|=9/4$)
radicals, in the approximately 0.03~cm$^3$ detection volume in
the trap. In the inset, the LIF signal intensity is shown on a
longer time-scale, from which a 1/e trap lifetime on the order of
1.0~second, limited by collisions with background gas, is deduced.

\begin{figure}
    \centering
    \resizebox{\linewidth}{!}
    {\includegraphics[64,366][343,623]{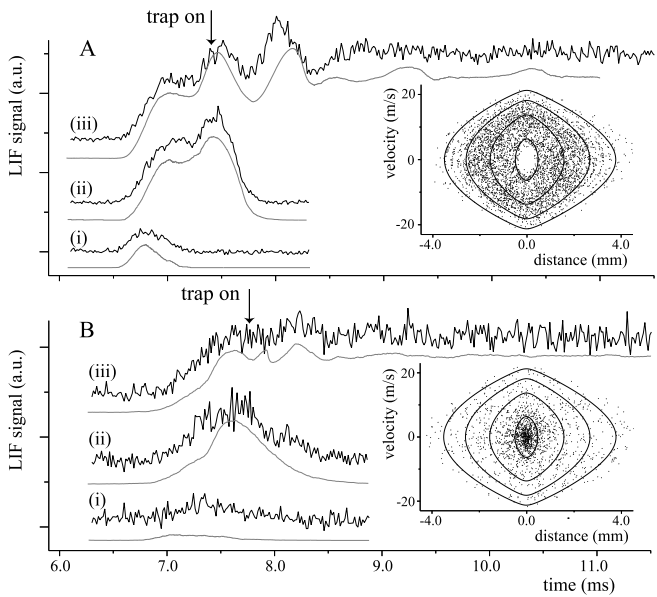}}
    \caption{LIF signal of OH ($J=3/2$) radicals at the center of the trap as a
    function of time after OH production for two different trap loading
    sequences (case A and B), together with the numerically simulated curves. The calculated
    longitudinal phase space distribution of the OH radicals in the trap at $t$ = 30~ms is
    shown for both cases.}
    \label{fig:Trap-loading}
\end{figure}

The absolute number of OH radicals that can be trapped as well as
the phase space distribution of the trapped molecules critically
depends on the details of the trap loading sequence. In
Fig.~\ref{fig:Trap-loading} measurements are shown for two
slightly different trap loading conditions, referred to as case A
and case B. Curve A(i) and B(i) show the TOF profile of the slow
beam of OH radicals that passes through the trap when no voltages
are applied to the trap electrodes. Due to the lower beam velocity
of around 15~m/s in case B compared to the 21~m/s in case A, the
beam arrives later at the center of the trap, is spread out more
and has a lower peak density. Curve A(ii) and B(ii) show the TOF
profile when the voltages for trap loading as given earlier are
constantly applied to the trap electrodes. In case A the
molecules enter the trap too fast, and move past the center of
the trap; the observed double-hump structure in the TOF profile
is to be interpreted as the signal from both the incoming and the
reflected beam. In case B the molecules are reflected almost
exactly at the center of the trap. Curve A(iii) (identical to the
lower curve in Fig.~\ref{fig:TOF}) and B(iii) show the resulting
TOF profile when we switch from the loading geometry to the
trapping geometry at the time indicated by the arrow. The higher
initial velocity in case A leads to approximately a factor three
more molecules in the trap, but also results in clear
oscillations in the TOF profile. When the molecules are (more)
correctly coupled into the trap, as in case B, hardly any
oscillations are observed.  The simulations underneath the
experimental data accurately reproduce the observed TOF profiles.
The calculated longitudinal phase space distribution of the
molecules in the trap at $t$ = 30~ms, shown in
Fig.~\ref{fig:Trap-loading} for both cases, should therefore also
be realistic. From this, the FWHM of the velocity distribution of
the trapped molecules is deduced to be around 30~m/s in case A
and around 10~m/s in case B, corresponding to a temperature of
450~mK and 50~mK for the trapped OH radicals, respectively. The
density is rather similar in both cases, and is about 10$^7$
cm$^{-3}$.

The samples of electrostatically trapped OH radicals produced
here, are, for instance, ideally suited to investigate the
predicted linking of ultracold polar molecules
\cite{Avdeenkov2003}. The experiments reported here are performed
in a molecular beam deceleration and trapping machine that is
designed such that almost the complete molecular beam pulse that
passes through the skimmer in the right quantum-state can be
transported through the trap at a tunable absolute velocity while
staying together as a compact package. Alternatively, a sizeable
fraction of this molecular beam pulse can be slowed down to a near
standstill and loaded into the electrostatic trap, where the
molecules can then be trapped up to seconds. Sensitive
quantum-state specific detection of the molecules can be
performed inside the quadrupole trap. As the molecular beam
machine typically runs at a 10~Hz repetition rate, and as
different high voltage switching sequences can be applied to
adjacent molecular beam pulses, this together offers the unique
possibility to perform "in-beam" collision and reactive
scattering experiments as a function of the continuously tunable
collision energy and with an unprecedented energy resolution.

\begin{acknowledgments}
This work is part of the research program of the `Stichting voor
Fundamenteel Onderzoek der Materie (FOM)', which is financially
supported by the `Nederlandse Organisatie voor Wetenschappelijk
Onderzoek (NWO)'. This work is supported by the EU "Cold Molecules"
network and by a fellowship (for R.T.J.) of the Royal Netherlands
Academy of Arts and Sciences.
\end{acknowledgments}

\end{document}